\documentclass[prb,showpacs,preprint]{revtex4}
\usepackage {graphicx}

\begin{document}

\title{Conductivity landscape of highly oriented pyrolytic graphite surface containing ribbons and edges}

\author{S. Banerjee\footnote{On Deputation from Surface Physics Division, Saha Insitute of Nuclear Physics, 1/AF Bidhannagar, Kolkata,{\bf email:}sangam@hp1.saha.ernet.in}, M. Sardar, N. Gayathri, A. K. Tyagi and Baldev Raj}

\affiliation{Materials Science Division,\\ Indira Gandhi Centre for Atomic Research, \\
Kalpakkam – 603102, T.N, India}

\begin{abstract}
We present an extensive study on electrical spectroscopy of graphene ribbons and edges of highly oriented pyrolytic graphite (HOPG) using atomic force microscope (AFM). We have addressed in the present study two main issues, (1) How does the electrical property of the graphite (graphene) sheet change when the graphite layer is displaced by shear forces? and (2) How does the electrical property of the graphite sheet change across a step edge? While addressing these two issues we observed, (1) variation of conductance among the graphite ribbons on the surface of HOPG. The top layer always exhibits more conductance than the lower layers, (2) two different monolayer ribbons on the same sheet of graphite shows different conductance, (3) certain ribbon/sheet edges show sharp rise in current, (4) certain ribbons/sheets on the same edge shows both presence and absense of the sharp rise in the current, (5) some lower layers at the interface near a step edge shows a strange dip in the current/conductance (depletion of charge). We discuss possible reasons for such rich conducting landscape on the surface of graphite.

\end{abstract}

\pacs{}

\maketitle

\section{Introduction}
Extensive theoretical and experimental studies has been made on layered graphite material to understand its complex electrical behaviour \cite{Walker,Setton}. Recently, electronic properties of graphene nano-structures, such as carbon nanotubes, carbon nanocones, fullerenes and graphite edges, have attracted much attention from the point of view of basic research and applications. The motivation for studying carbon based nano-structures is to develop atomic or molecular nanometer scale electronic devices having fundamentally different operating principle than conventional electronic devices. Due to its nanoscale size one can look at opportunities to increase the device density as well. The electronic property of nanoscale materials are strongly influenced by their geometries. The graphene sheets are generally self-assembled to arrive at various shaped nano-objects such as carbon single-walled and multi-walled nanotubes, fullerenes, nanocones etc., \cite{Setton,Popov,Jordan,Saito}. The electronic properties of these objects strongly depend on the manner in which these graphene sheets are cut and joined at the edges. Any new information and understanding related to graphene sheets will have an impact on basic research and applications of carbon based materials.

Highly oriented pyrolytic graphite (HOPG) is a periodical stack of two-dimensional (2D) graphene sheets (layers) along c-axis. Each sheet comprises of hexagonal lattice of carbon bonded by strong $\sigma$ bonding ($sp^2$) in the a-b plane (see fig. 1(a)). The perpendicular (along c-axis) $\pi$-orbital electrons are responsible for the conductivity along the a-b plane. The conduction occurs by the quantum mechanical hopping of these electrons. Each of these layers are weakly bonded to their neighbouring layers by interlayer interaction forces. Because of the weak interlayer interaction forces, the graphene layers can easily slide against each other and peel off easily. In late 50's and early 60's \cite{Slonczewshi,McClure} the electronic property of graphite was evaluated using phenomenological models based on the symmetry of graphite. The dispersion relation of the bands was found to have a very slight overlap of the valence and conduction bands ($\sim$ 0.04 eV) at the Fermi level, where the electron density is very low. This causes graphite to have semi-metallic characteristics. Eventhough the charge carrier concentration is very low ($\sim 10^{-4}$ per C atom), the electrical mobility of these carriers ($\sim 10^{4}$ to $10^{5}$ $cm^2V^{-1}s^{-1})$ is high and its electrical resistivity along the plane ($\rho_\parallel$) is $\sim 40~\mu \Omega cm$ \cite{Walker,Setton1}. The temperature coefficient of resistivity of HOPG is positive (metallic) along the sheet and $\rho_\perp/\rho_\parallel \approx 10^4$ at room temperature. 

It has been pointed out that graphene sheet edges strongly affect the $\pi$ electronic states 
\cite{Fujita}. The edges of the graphene sheets are of two types (1) armchair (cis) and (2) zigzag (trans) edges. These two edges are shown in Fig. 1(b) and Fig. 1(c) respectively. It was shown theoretically that graphene sheets having zigzag edges possess edge states localized \cite{Fujita,Nakada,Fujita1,Stein,Hosoya,Tanaka,Fujita2,Ryu,Wakabayashi,Elstner} at the zigzag edges. In contrast, armchair edges have no edge states at its edges, hence making the armchair edge less conducting than the zigzag edge. 

With this introduction on graphite, we would like to address two main issues in the present paper: 

1. How does the electrical property of the graphite (graphene) sheet change when the graphite layer is displaced by shear forces?

2. How does the electrical property of the graphite (graphene) sheet change across a step edge? 

STM and AFM with conducting tip can be used to study the local electronic properties of conducting surfaces. Using STM measurement one can obtain images and perform spectroscopy with very high spatial resolution. In the STM one adopts either a constant height mode or a constant current mode for mapping. Mapping conductivity across step edges using STM has some difficulties. In the constant height mode, the variation of current measured across the step is due mainly to the height variation and hence mapping local conductivity using constant height mode is impractical. In the constant current mode, since the current is kept constant it is difficult to map the local conductivity from the displacement of the z-piezo because tunnelling current decreases exponentially with distance, thus small current change from sheet to sheet will lead to very small displacement of the z piezo. But, since edge states have larger conductance (as will be seen below), STM in the constant current mode can be used to see these edge states. Performing I-V measurements at every point by stopping the feed-back signal to map the local conductivity with high resolution is also difficult using STM. AFM with a conducting tip in the contact mode overcomes the above difficulties at the expense of spatial resolution. For the present investigation we have used AFM in contact mode with constant applied normal force on the tip.

The paper is organized as follows. Experimental details are given in section II. In section III, the experimental results are presented. Section IV contains the discussion of the results.

\section{Experimental details}
For the present investigation we have used freshly cleaved highly oriented pyrolytic graphite HOPG. By (simply) peeling off the surface layers using scotch tape, we were able to obtain ribbons and terraces with step edges. The peeling off process dislocates the layers laterally and vertically. We have studied only monolayer steps to avoid complexity in the electrical analysis. Commercial AFM (NT-MDT, Russia) was used for the present investigation. Cantilevers used were platinum coated (CSG 10/Pt, NT-MDT) with radius of curvature of the tip $\sim$ 35 nm and the cantilever elastic constant 0.1 N/m. For imaging in the the contact mode the normal force between the surface and the tip was kept around $\sim$ 25 nN (constant force mode). The sample was biased with 10 to 15mV for measuring the local conductance (spreading resistance imaging) of the sample surface with the conducting tip. By measuring the tip-sample contact conductance (resistance) during the scanning process, we measure the local conductance of the sample surface. For purely Ohmic contacts (i.e., metal-metal contacts), the I-V is of the typical spreading resistance form:

\begin{equation}
R_{spreading}=\frac{\rho}{2a}
\end{equation}

\noindent where $\rho$ is the mean resistivity of the tip-sample contact and $a$ is the contact area. All the measurements were conducted at room temperature and ambient conditions. For topographic images we have only subtracted linear line fit along the scan direction from the scanned images and presented without any further filteration or averaging. For the conductance map we show only the raw results without any processing (processing the data was avoided as this can sometime be misleading).

\section{Experimental results}

In fig. 2(a,b) we show the topographic image and local conductance (spreading resistance) map of a freshly cleaved HOPG sample. We see that the contrast in the local conductance map is much better than the topographic image. Each seperate graphite ribbons can be easily distinguished in the local conductance map. The local conductance image (fig. 2b) of the HOPG sample clearly indicates that the conductivity of these ribbons are different since the conductance was mapped in a constant force mode (which ensures that the contact area remains constant during the whole scan over the sample surface). Hence the tip-sample contact conductivity measured, is a direct measure of the local conductivity of the sample and is not due to change in the area of the contact (see eqn.~1). Since we have used a 35nm radius of curvature tip, we can safely say that the tip-sample contact area is limited by this and will be less than this value. 

In fig. 3(a,b) we show topographic two-dimensional (2D) and three-dimensional (3D) images  of a selected region containing two monolayer steps/edges (staircase like) on the surface of the HOPG sample. In fig. 3(c) we show two line profiles marked in fig. 3(a) as A and B. We can clearly see the step heights to be around $\sim$0.4 nm close to the reported c-axis value of 0.35 nm for the HOPG sample (The slightly higher value obtained will be explained in section IV). In fig. 4(a,c) we show the local conductance map of these steps for the forward and the reverse biased condition respectively. We can clearly see the contrast inversion and reversal of the current direction on reversing the bias. The higher current is indicated by brighter shade and the lower current by darker shade in the forward bias condition and the scale is shown on the right hand side of the image. In the reverse bias condition the higher negative current value is indicated by darker shade and low negative value is shown as brighter shade. In fig. 4(a) the brightest layer is the top layer and the darkest is the lowest layer. Detail discussion on the difference in the current contrast will be discussed in the next section. In fig. 4(b,d) we show the line profile of the current along the lines marked in the conductance images (fig. 4(a,c) respectively). We see distinct changes (drop) in the current profile at the step edges in comparison to the height change in the topography (see fig. 3(c)). This clearly demonstrates why the contrast is better in the conductivity map than the topographic map. At a closer look we see distinct dips in the current occuring at the lower layers at each of the interface of the step edges (marked by arrows in fig. 4(b,d)). Since the conductance depends on the local electron density, the dip in the conductance of the lower layers near the step edge is the manifestation of the depletion of the local electron density. This feature has never been observed earlier to our knowledge using either STM or AFM. 

In fig. 5(a,b) insets we show two conductance maps taken at two different regions showing bright current streaks appearing on certain edges as indicated by arrows. This observation indicates clearly the presence of two distinct type of edges on this graphite ribbons. This is more illustrative in the line profile of the conductance current shown in fig. 5(a,b) of the lines marked A, B, C and D in fig. 5(a,b) insets. The current at the edges showing the bright streaks are 2 to 4 times the value measured on the terrace. In this case, the lower layer near the step edge does not show a dip in the conductance/current as observed earlier (fig. 4). It appears that the high electron density at the edge of the top layer has compensated for the deficient electron density of the below layer near the top layer edge. 

In fig. 6(a,b) we show topographic 2D and 3D images of a well-shaped monolayer step region. This is more clearly seen from the line profile shown in fig. 6(c) for the lines marked in fig. 6(a). The region I and II are of the same height but the conductance are different. In fig. 7(a,c) we show the local conductance map for the forward and reverse bias condition. We see a brighter shade for region I (indicating higher conductance) than for region II for the forward bias condition, eventhough both the sheets are monolayers and on top of the same sheet III. This may imply that the interlayer bonding for layer I and layer II with III are different. The current contrast gets reversed on reversing the bias (fig. 7(b)). Another interesting feature we observe is that the bright streak abruptly ends along the edge (dark streak in the case of reverse bias) of the same ribbon as shown by arrows in fig. 7(a,c). (Note: The images of fig. 7(a) and (c) are shifted due to thermal drift which was unavoidable with the present commercial system). This indicates that the  edge can be of two different types on the same side of a single ribbon/sheet. In fig. 7(b,d) we show the current profile along the lines marked in fig. 7(a,c) for forward bias condition and the reverse bias conditions respectively. 

We have also performed current-voltage (I-V) measurement on these sheets as shown in fig. 8(a, b) at various regions of ribbons/terraces marked in fig.~5(b) inset and 7(a) respectively. We clearly see that the ribbons which are bright in the forward bias condition in the local conductance images show large slope in the I-V measurement at zero bias. The intermediate conductance ribbons (according to the grey scale of the conductance image) show intermediate slopes and the least conductance ribbons show the lowest slope value. As we have pointed out earlier, the normal force $F_N$ of the tip on the sample was kept constant during the imaging and during the I-V measurements and hence the measured conductance is only due to the local conductivity. Thus the conductance map and the I-V measurements indicates that the ribbons/sheets which are dislogded or sheared have more conductivity than the sheets which are held with the bulk.

We now summarise the main observations of this study:

(1) variation of conductance among the graphite ribbons on the surface of HOPG. The top layer always exhibits more conductance than the lower layers (figs. 4,5 and 7).

(2) two different monolayer ribbons on top of the same graphite sheet shows different conductance (figs.~6 and 7).

(3) certain ribbon/sheet edges show sharp rise in current (figs.~5 and 7).

(4) certain ribbons/sheets on the same edge shows both presence and absense of the sharp rise in the current (fig.~7).

(5) some lower layer at the interface near a step edge shows a strange dip in the current conductance (depletion of charge fig.~4 ). 

We will discuss the implications of the above observations in the next section.

\section{Discussion}
The first observation is the variation of conductance of graphite ribbons on the surface of HOPG. The top layer always exhibits more conductance than the lower layers (figs.~. 4, 5 and 7). During the peeling off process the top few layers of graphite peels off inhomogeneously forming ribbon like structure as shown in fig. 2. The ribbon edges formed will have monolayered and multilayered steps at the edges. So, during the peeling off process it will exert vertical and lateral stresses on the ribbons. Due to applied vertical and lateral stresses during the pulling process, these ribbons will get dislocated vertically and laterally. It is known that $\pi$-orbitals which are perpendicular to the graphite sheet are responsible for the electrical conductance along the sheet.  It is also believed that the interlayer forces are the van der Waals' forces. We believe that the $\pi$ electrons not only take parts in conductivity, but also have to contribute substantially to the polarisation cloud (without electron transfer) that gives the bonding between the layers. If the top layer is loosely held, then the $\pi$-electrons of the loosly held top layers donot participate much in the bonding with the layer underneath. This makes the $\pi$ electrons more mobile leading to higher conductivity. Note: The carrier density does not change and it is only the mobility which increases thus leading to a higher conductivity.

Another important way in which the top layer may be more conducting could be due to an unavoidable crumpling of the top graphite sheet (small displacement of carbon atoms in and out of the plane) especially whenever such distortions are unrelaxed. This effectively introduces the next nearest neighbour hopping amplitude if one thinks of the $\pi$ electrons in terms of the tight binding approximation. The six isolated points in the Brillouin Zone (BZ) where the valence band (VB) and the conduction band (CB) touches for only nearest neighbour hopping case is shown in fig.~1(d). With the introduction of the next nearest neighbour hopping (induced due to the displacement of the carbon atoms out of the plane due to crumpling) expands these isolated points to a finite area as shown in fig.~1(e). This means larger density of low energy current carrying states at Fermi level, resulting in an increased conductivity. In other words, likeliness of having local unrelaxed crumpling of sheets is larger for the top layers and any crumpling automatically dopes the zero-gap semi-metal with added carriers. 

The peeling of the layers were performed manually in an uncontrolled fashion. This can lead to different degrees of applied stress on different ribbons and cause different magnitude in the dislodgement. It can happen that two ribbons seperated by distances but lying on top of the same graphite sheet can be dislodged in different proportion i.e., the interlayer c-axis distance can be different. The one which has a larger c-axis distance will be loosely held than the ribbon which has a lower c-axis value (fig.~7). The difference in conductance can be explained invoking the same arguments given for the first observation i.e, the loosely held ribbons are more conducting than the tightly held layers. Thus in fig.~7 we observe that the two adjacent layers having almost a similar height seperated by distance show different conductance. We can infer that the layers which show more conductance is less tightly bond to the layers underneath or the layers might be crumpled. With our resolution we cannot check the crumpling of top layers. It needs more careful and higher resolution studies.

Now, we will discuss the presence of sharp current peaks and dips at the edges. In fig.~9(a,b,c) we motivate schematically the step edges of different kind that can give rise to the corresponding current versus distance (x) profile across the step edges and layers. Certain sheets/edges show a sharp rise in the conductance (fig.~5 and 7). As mentioned in the introduction, there are two types of edges formed on graphene sheets namely zigzag and armchair. During the tearing process of the graphite sheets to produce ribbons and steps, the sheets are torn forming these two edges. Earlier detailed theoretical calculations have shows that zigzag edges have edge states localised at the edges having a higher electron density \cite{Fujita,Nakada,Fujita1,Stein,Hosoya,Tanaka,Fujita2,Ryu}. Hence these edges will have higher electrical conductance. Thus during the mapping of the local conductance these edges will show up as brighter streaks (indicating more conductance). During the tearing process there is equal probability that the edges formed will be of either zigzag type or armchair type. The same edge can contain both these type of edge shapes. This can also happen due to the mosaicity of the graphite sheet. Each mosaic can be oriented in different direction and during the tearing process it will tear along any suitable (either zigzag or armchair) direction. As mentioned earlier the armchair edges have no edge states and are thus less conducting compared to the zigzag edges. It is not clear whether reconstruction is taking place at the edges. We can only say that the appearance of bright streak or dissapearance of it depends on the the shape of the edge.

The peculiar dip in current (fig.~4 and fig.~9(a)) observed at some edges is puzzling i.e., the formation of a charge depleted region. As we have argued earlier, the interlayer bonding utilises the $\pi$-orbitals to a certain extent. Near the step edge if there are no excess charges (no sharp peak) then, it seems like that the electrons from the $\pi$-orbitals of the underlying layer have been dragged from the vicinity of the edge causing depletion of charge in that region. If the ribbons are very loosely held then this effect diminishes or not observed. This observation was not reported earlier and will need more theoretical understanding of the role of $\pi$-orbitals in the interlayer coupling apart from its role in electrical conduction. 

\section{Summary}
We have demonstrated that AFM with conducting tip can also be used and in certain cases may produce better results than STM in the measurement of local conductivity mapping at the cost of spatial resolution. From the present study we could say that $\pi$-orbitals may play an important role in the interlayer forces apart from electrical conductance. When the top layers of the graphite sheets are loosely held with the bulk graphite then the conductivity of the layer increases.  The increase in the conductivity of these sheets can be explained only by assuming that, the $\pi$ electrons on the loosely held sheets, are not much involved in binding between the layers and hence would have higher mobility (conductivity). Additionally, crumpling of top layers where ever present will also create additional carriers in the plane. Sharp increase or absence in conductance current at the edges of the graphite sheets have been attributed to either zigzag shape edges or armchair shape edges respectively. We have also observed the presence of these two types of edges on a single edge of graphite sheet. For the first time a charge depleted region near the step edge on the lower layer graphite sheet is reported. We think that all the above experimental observations are very important and will require more theoretical understanding and will have new impact on the understanding of the graphene layers.

\newpage
$\bf{References:}$

\newpage

\noindent $\bf{Figures Captions}$

\noindent $\bf{Fig. 1}$: (a) Carbon atoms showing $sp^2$ hybridized $\sigma$ orbitals in the a-b plane and nonhybridized $\pi$ orbitals along the c-axis of graphite layer. Graphite sheet/ribbons showing (b) armchair (cis) edge and (c) zigzag (trans) edge. (d) Six points in the Brillouin zone showing zero-gap and having zero-overlap of the conduction band (CB) and the valence band (VB) at Fermi energy ($E_F$) i.e., zero density of state (DOS) at $E_F$ and (e) finite overlap of CB and VB leading to finite DOS at $E_F$. 

\noindent $\bf{Fig. 2}$: Atomic force microscopic image (a) topographic (b) conductivity landscape of highly oriented pyrolytic graphite (HOPG).

\noindent $\bf{Fig. 3}$: (a) Topographic two-dimensional (2D) and (b) three-dimensional (3D) images  of a selected region containing two monolayer steps/edges (staircase like) on the surface of the HOPG sample. (c) Two height profiles along the lines marked in (a) as A and B. The arrows indicate the 0.4 nm step height. (Note: The tilt in the step (line profile) is an artifact of processing (linear line fit along the scan x-direction))

\noindent $\bf{Fig. 4}$: (a) Local conductance map of steps of fig.~3 for the forward and (c) for the reverse bias conditions respectively. Contrast inversion and reversal of the current direction is observed on reversing the bias. The higher current is indicated by bright shade and lower current by darker shade in the forward bias condition and the scale is shown on right hand side of the image. In the reverse bias condition the higher negative current value is indicated by darker shade and low negative value is shown as brighter shade. The brightest layer in (a) is the top layer and the darkest is the lowest layer. (b) and (d) current profile along the lines marked in the conductance images (a and c) respectively. We see distinct changes (drop) in the current profile at the step edges in comparison to the height change in the topography (see fig.~ 3(c)). The arrows show distinct dips in the current/conductance occuring at the lower layers at each of the interface of the step edges.

\noindent $\bf{Fig. 5}$: (a and b) insets show two conductance maps taken at two different regions showing bright current streaks appearing on certain edges as indicated by arrows. This observation indicates clearly the presence of two distinct type of edges on this graphite ribbons. (a) and (b) show line profiles of the conductance current of the lines marked A, B, C and D respectively in insets.

\noindent $\bf{Fig. 6}$: (a) shows topographic 2D and (b) 3D images of a well-shaped monolayer step region. This is more clearly seen from the height profile shown in (c) for the lines marked in (a) as A and B. The region I, II and III are the three regions explained in the text. 

\noindent $\bf{Fig. 7}$: Local conductance map for the (a) forward and (c) reverse bias condition. I, II and III are the three different regions. labels 1 to 8 are the spots where current versus voltage (I-V) measurements were carried out.  (b) The current profile along the lines marked in (a) with label A to E for forward bias condition and (d) for the reverse bias conditions for lines F to K respectively. The arrows indicate the bright streak abruptly ends along the edge for the forward bias condition and the end of dark streak in the case of reverse bias of the same ribbon. (Note: The images of (a) and (c) are shifted due to thermal drift which was unavoidable with the present commercial system). 

\noindent $\bf{Fig. 8}$: (a) and (b) Current-voltage (I-V) measurement carried out at various regions of ribbons/terraces marked in fig.~5(b) inset and 7(a) respectively. Low slope value in the I-V curve at zero-bias indicates lower conducting region than the region having higher slope. (Note: all the I-V measurement was carried at same value of constant force).

\noindent $\bf{Fig. 9}$: In (a), (b) and (c) we show schematically the step edges of different kind that can give rise to the corresponding current versus distance (x) profile accross the step edges and layers.

\begin{figure}
\centerline{\includegraphics*{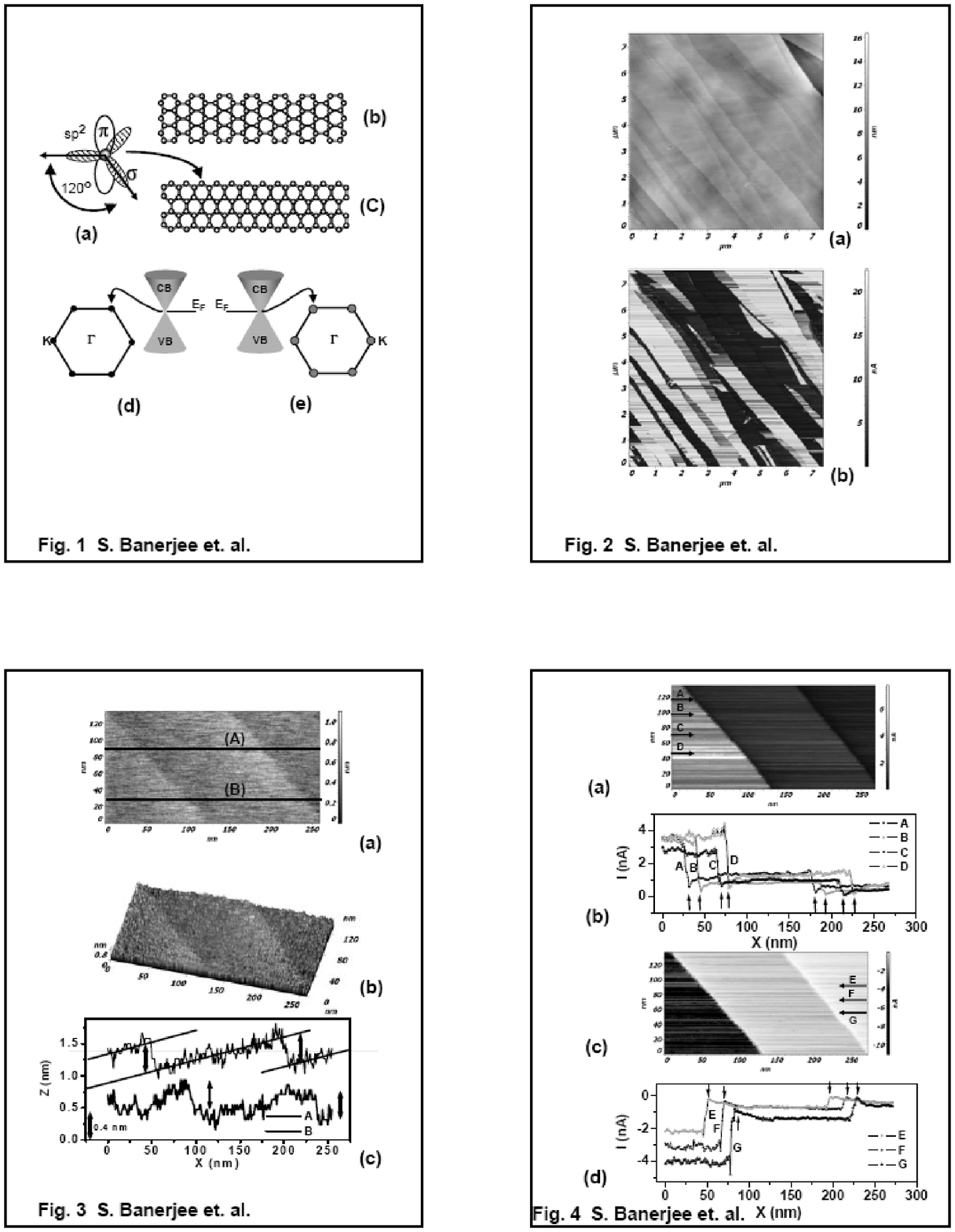}}
\end{figure}

\begin{figure}
\centerline{\includegraphics*{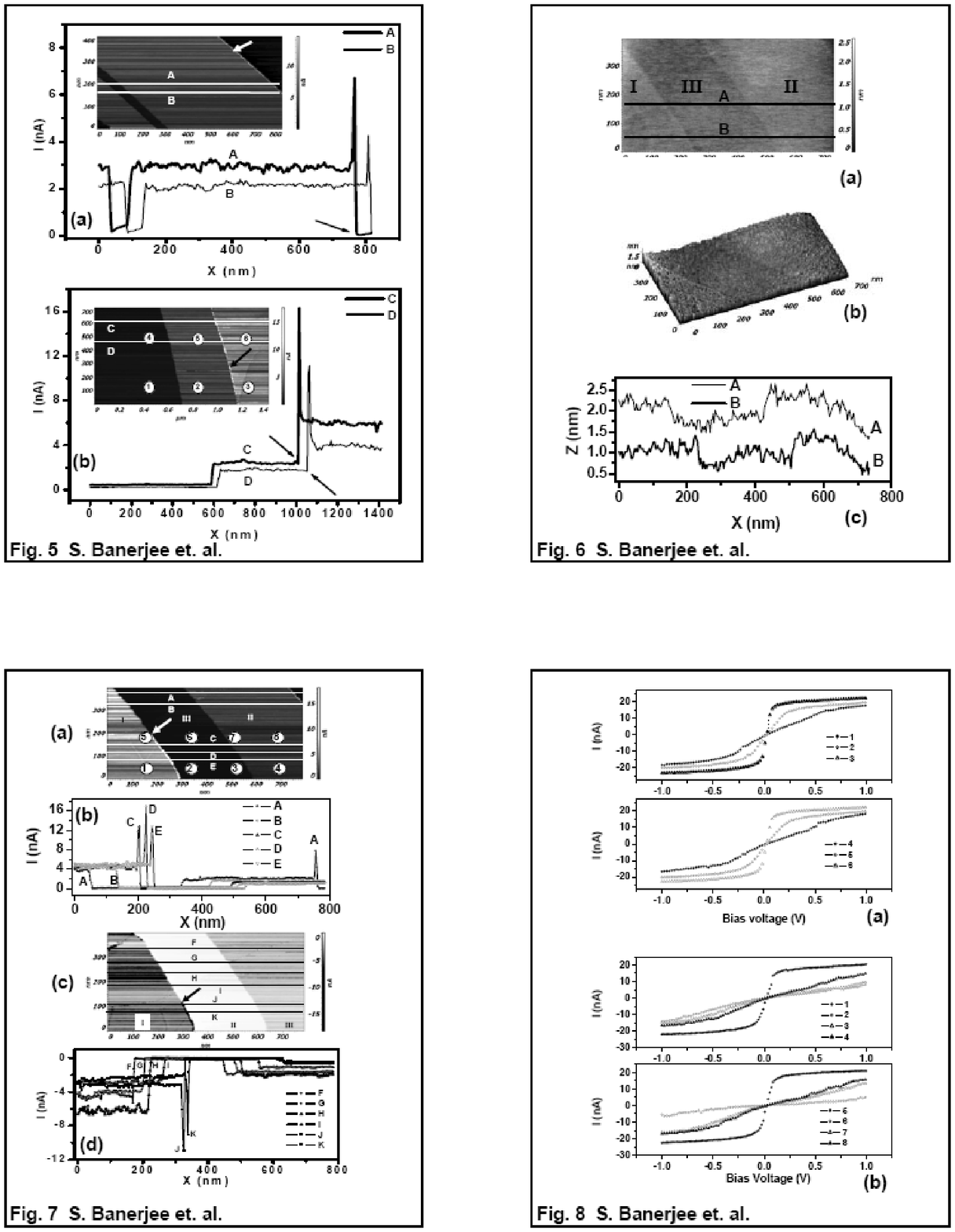}}
\end{figure}

\begin{figure}
\centerline{\includegraphics*{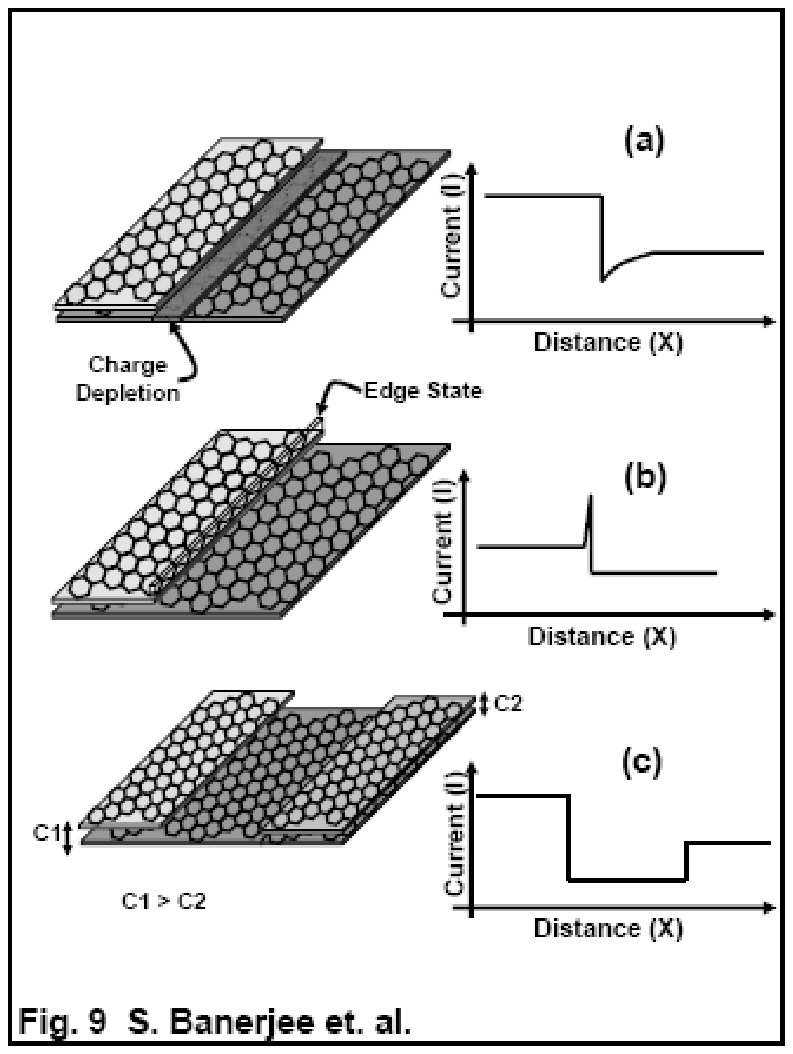}}
\end{figure}

\end{document}